\documentclass[a4paper,12pt]{article}
\usepackage{amssymb}
\usepackage{graphicx}

\def\grad{\vec{\triangledown}}
\def\o{\vec{\omega}}

\def\l{{\cal L}}

\def\pd{\partial}
\def\d{\mathrm{d}}

\def\jt{\tilde{J}}
\def\jtv{\vec{\tilde{J}}}

\def\be{\begin{equation}}
\def\ee{\end{equation}}
\def\bea{\begin{eqnarray}}
\def\eea{\end{eqnarray}}
\def\ie{\textit{i.e.} }
                                 
\def\ie{\textit{i.e.} }
\def\pd{\partial}
\def\grad{\vec{\triangledown}}
\def\l{{\cal L}}

\def\ja{{\cal J}}
\def\A{{\cal A}}
\def\D{{\cal D}}
\def\F{{\cal F}}
\def\o{\vec{\omega}}

\def\uv{\vec{v}}
\def\av{\vec{A}}

\def\pd{\partial}
\def\d{\mathrm{d}}

\def\jt{\tilde{J}}
\def\jtv{\vec{\tilde{J}}}

\def\be{\begin{equation}}
\def\ee{\end{equation}}
\def\bear{\begin{eqnarray}}
\def\eear{\end{eqnarray}}

\def\({\left(}
\def\){\right)}
\def\[{\left[}
\def\]{\right]}

.360pk

\title{Lagrangian dynamics of the Navier-Stokes equation}

\author{A. Sulaiman$^{\mathrm{a,c}}$\footnote{Email : lyman@tisda.org} 
\, \, and \, \, 
L.T. Handoko$^{\mathrm{b,c}}$\footnote{Email : handoko@fisika.lipi.go.id, handoko@fisika.ui.ac.id}}
\date{}

\begin{document}

\maketitle

\thispagestyle{empty}

\begin{center}
\begin{small}
\noindent
$^\mathrm{a)}$ Geostech BPPT\footnote{http://www.bppt.go.id}, Kompleks Puspiptek Serpong, Tangerang 15310, Indonesia \\
\vspace*{2mm} 
$^\mathrm{b)}$ Group for Theoretical and Computational Physics, Research Center for Physics, Indonesian Institute of Sciences\footnote{http://www.fisika.lipi.go.id}, Kompleks Puspiptek Serpong, Tangerang 15310, Indonesia \\
\vspace*{2mm} 
$^\mathrm{c)}$ Department of Physics, University of Indonesia\footnote{http://www.fisika.ui.ac.id}, Kampus UI Depok, Depok 16424, Indonesia \\
\end{small}
\end{center}

\vspace*{5mm}

\begin{abstract}
Most researches on fluid dynamics are mostly dedicated to obtain the solutions of Navier-Stokes equation which governs fluid flow with particular boundary conditions and approximations. We propose an alternative approach to deal with fluid dynamics using the lagrangian. We attempt to develop a gauge invariant lagrangian which reconstructs the Navier-Stokes equation through the Euler-Lagrange equation. The lagrangian consists of gauge boson field $\A_\mu$ with appropriate content describing the fluid dynamics, \ie  $\A_\mu = (\Phi, -\uv)$. An example of applying the lagrangian to the interaction of fluid in a solitonic medium is also given.
\end{abstract}

\clearpage

\section{Introduction}

The fluid dynamics still remains as an unsolved problem.
Mathematically, a fluid flow is described by the Navier-Stokes
(NS) equation \cite{kundu}:
\be
\frac{\pd\uv}{\pd t} + (\uv\cdot\grad)\uv = -\frac{1}{\rho}\grad P
- \mu \grad^2 \uv \; , 
	\label{eq:naviers} 
\ee 
where $\uv$ is fluid velocity, $P$ is pressure, $\rho$ is density and 
$\mu$ is the coefficient of viscosity.\\

In principle, the study of fluid dynamics is focused on solving the
Navier-Stokes equation with particular boundary conditions and / or 
some approximations depend on the phenomenon under consideration. 
Mathematically it has been known as the boundary value problem.
The most difficult problem in fluid dynamics is turbulence
phenomenon. In the turbulence regime, the solution for the
Navier-Stoke equation has a lot of Fourier modes, such that the
solution is untrackable numerically or analytically. It is
predicted that the strong turbulence has $10^{10}$ numerical
operation \cite{mulin}. This motivates us to look for another approach 
rather than the conventional ones. This paper treats the fluid
dynamics differently than the conventional point of view as seen
in some fluid dynamics textbooks. In this approach, the fluid is
described as a field of fluid buch. We use the gauge field theory
to construct a lagrangian describing fluid dynamics by borrowing the
gauge principle. The Navier-Stoke
equation can be obtained from this Lagrangian as its equation of
motion through the Euler-Lagrange principles.

\section{Maxwell-like equation for ideal fluid}

The abelian gauge theory $U(1)$ is an electromagnetic theory that
reproduces the Maxwell equation. To build a lagrangian that is
similar with the abelian gauge theory, we should 'derive' the
Maxwell-like equation from the Navier-Stokes equation \cite{sulaiman}. 
The result
can be used as a clue to construct a lagrangianan for fluid that
satisfies gauge principle. Considering the Navier-Stokes equation
Eq. (\ref{eq:naviers}) for an ideal and incompressible fluid, 
\bear
 \rho \left( \frac{\pd \uv}{\pd t}+(\uv.\grad)\uv \right) & = & -\grad P  \; , 
 \label{eq:NSideal1} \\
 \grad\cdot\uv & = & 0 \; .
 \label{incompres}
\eear 
Using the identity
$\uv\times(\grad\times \uv)=\grad(\frac{1}{2}\uv^{2})-(\uv\cdot\grad)\uv$,
it can be rewritten as,
\be
 \frac{\pd \uv}{\pd t}+\grad \left( \frac{1}{2}\uv^{2} \right) -\uv\times(\grad\times\uv)
 = -\frac{1}{\rho}\grad P \; ,
\ee and then,
\be
 \frac{\pd \uv}{\pd t} = \uv\times(\grad\times\uv)
 -\grad \left( \frac{1}{2}\uv^{2}+\frac{P}{\rho} \right) \; .
\label{eq:NSideal2} 
\ee 
Putting the scalar potenstial
$\Phi=\frac{1}{2}\uv^{2}+\frac{P}{\rho}$ , the vorticity
$\vec{\omega}=\grad\times\uv$ and the Lamb's vector
$\vec{l}=\vec{\omega}\times\uv$, the equation becomes, 
\bear
 \frac{\pd \uv}{\pd t} & = & -\vec{\omega}\times\uv-\grad\Phi
 \nonumber \\
 \label{eq:NSideal3}
 & = & -\vec{l}-\grad\Phi \; .
 \label{eq:NSideal4}
\eear
Imposing curl operation in Eq. (\ref{eq:NSideal3}) we obtain the
vorticity equation as follow,
\be
 \frac{\pd \vec{\omega}}{\pd t}=-\grad\times(\vec{\omega}\times\uv)\; .
 \label{eq:vortisitas}
\ee \\

In order to get the Maxwell-like equation for an ideal fluid, let us
take divergence operation for Eq. (\ref{eq:NSideal4}), that is 
\bear
 \frac{\pd}{\pd t}(\grad\cdot\uv)&=&-\grad\cdot\vec{l}-\grad^{2}\Phi\nonumber \\
 \grad\cdot\vec{l}=-\grad^{2}\Phi &=& \tilde{\rho} \; .
 \label{eq:divlamb}
\eear 
Here we have used the incompressible condition, while by definition the 
divergence of vorticity is always zero, \ie 
$\grad\cdot\vec{\omega}=0$. Imposing again curl operation, we have, 
\bear
 \frac{\pd}{\pd t}(\grad\times\uv) & = & -\grad\times\vec{l}-\grad\times(\grad\Phi) \; , \nonumber \\
 \frac{\pd\vec{\omega}}{\pd t} &=& -\grad\times\vec{l} \; , \\
 \grad\times\vec{l} &=& -\frac{\pd\vec{\omega}}{\pd t} \; ,
 \label{eq:curllamb} \nonumber
\eear 
using the identity $\grad\times(\grad\cdot\phi)=0$. \\

Now, let us consider the
definition of the Lamb's vector $\vec{l}=\vec{\omega}\times\uv$.
Taking the derivative ${\pd}/{\pd t}$ in the definition we obtain,
\be
 \frac{\pd\vec{l}}{\pd t} = \frac{\pd \vec{\omega}}{\pd t}\times\uv +
 \vec{\omega}\times\frac{\pd \uv}{\pd t} \; .
\label{eq:lamb1} 
\ee 
Substituting Eq. (\ref{eq:NSideal3}) and
(\ref{eq:vortisitas}), we get,
\be
 \grad\times\vec{\omega} = \alpha\vec{j}+\alpha\frac{\pd\vec{l}}{\pd t} \; ,
 \label{eq:curlvor}
\ee 
where, 
\bear
  \alpha & = & \frac{1}{\uv^{2}} \; ,
  \label{eq:alfa} \\
  \vec{j} & = & -\uv\grad^{2}\Phi + \left[
   \grad \times(\uv \cdot \vec{\omega}) \right] \uv
  +\vec{\omega} \times \grad(\Phi+\uv^{2})
  + 2 \left[ (\grad \times\uv) \cdot \grad \right] \uv \; .
  \label{eq:arus}
\eear 
These results induce a series of equations, 
\bear
 \grad \cdot \vec{l} &=& \tilde{\rho} \; ,
 \label{eq:divlamb1} \\
 \grad \times \vec{l}&=&-\frac{\pd \vec{\omega}}{\pd t} \; ,
 \label{eq:curllamb1} \\
 \grad \cdot \vec{\omega}&=&0 \; ,
 \label{eq:divvor1} \\
 \grad \times \vec{\omega}&=&\alpha\vec{j}+\alpha\frac{\pd\vec{l}}{\pd t} \; ,
 \label{eq:curlvor1} 
\eear
that is clearly the Maxwell-like equation for fluids.
If the fluid velocity is time independent, then $\vec{l}=-\grad\Phi$. 
This is the "electrostatic" condition. We use these results to develop gauge
field theory approach for fluid dynamics in the next section.

\section{Bosonic lagrangian for fluid}

The correspondences of the electromagnetism and the ideal fluid
can be written as follow, \bear
    \overrightarrow{B} & \leftrightarrow & \overrightarrow{\omega} \; , \nonumber \\
    \overrightarrow{E} & \leftrightarrow & \overrightarrow{l} \; ,\\
    \overrightarrow{A} & \leftrightarrow & \overrightarrow{\uv} \; , \nonumber \\
     \phi & \leftrightarrow & \Phi \nonumber \; ,
   \label{eq:corespondensi}
\eear where $\overrightarrow{B}$ is the magnetic field,
$\overrightarrow{E}$ is the electric field, $\overrightarrow{A}$
is the electromagnetics vector, $\phi$ is a scalar function,
$\overrightarrow{\omega}$ is the fluid vorticity,
$\overrightarrow{l}$ is the Lamb's vector, $\uv$ is fluid velocity
and $\Phi$ is the scalar potential. The same as the
electromagnetics field, we have a four vector $A_{\mu}=(\phi,\av)$
which can be interpreted as the four vector for fluid dynamics,
$ \A_{\mu} =(\Phi,-\uv)$. In the electromagnetics field, the scalar and 
vector potentials, $\phi$ and $\av$, are auxiliary fields. On the other hand, 
in the fluid dynamics the scalar potential $ \Phi= \frac{1}{2}
\uv^{2} + V$ describes the kinetic energy of fluid, while the vector 
potential $\uv$ is fluid velocity. Similar to the electromagnetics field,  the lagrangian density has the form of \cite{aslth},
\be
  \l_{NS} = -\frac{1}{4} F_{\mu\nu} F^{\mu\nu} + g \ja_\mu \A^\mu \; ,
  \label{eq:lagrange1}
\ee 
where,
\be
 \F_{\mu\nu}\equiv \pd_\mu\A_\nu - \pd_\nu \A_\mu \; .
 \label{eq:tensor1}
\ee
This Lagrangian obeys the gauge principles, \ie it is 
invariant under a particular local gauge transformation, 
\be
  \A_\mu  \rightarrow \A_\mu^{\prime} \equiv A_\mu + \frac{1}{g} \pd_\mu \theta \; ,
\label{eq:globalgauge} 
\ee 
where $\theta = \theta(x)$ is an arbitrary real constant. It is 
easy to show that the lagrangian density in Eq. (\ref{eq:lagrange1}) is
invariant under this transformation. \\

The equation of motion governed by this lagrangian can be derived using the 
Euler-lagrange equation in term of $\A_\mu$,
\be
  \pd^\nu \frac{\pd \l_{NS}}{\pd(\pd^\nu\A^\mu)}-\frac{\pd \l_{NS}}{\pd\A^\mu}=0 \; .
  \label{eq:EL}
\ee 
After a straightforward calculation, we obtain, 
\be
	\pd^\nu( \pd_\mu \A_\nu - \pd^\nu\pd_\nu \A_\mu)-g\ja_\mu = 0 \; .
	\label{eq:EL2} 
\ee
Now integrating it over $x^\nu$ and considering only the non-trivial 
relation as $\nu \neq \mu$ gives,
\be
  \pd_0 A_i - \pd_i A_0 = -g \oint \d x_0 J_i = g \oint \d x_i J_0 \; .
  \label{eq:EL4}
\ee Since $A_i=-\uv$, $A_o=\Phi$, $\pd_o= {\pd}/{\pd t}$ and
$\pd_i=\grad$. we have,
\be
-\frac{\pd\uv}{\pd t}-\grad\Phi=-g\jtv \label{eq:EL5}  \; , \ee
where $\jt_i \equiv \oint \d x_0 J_i = -\oint  \d x_i J_0$.
Concerning the scalar potential given by
$\Phi=\frac{1}{2}\uv^{2}+V$, we obtain,
\be
  -\frac{\pd \uv}{\pd t} - \frac{1}{2} \grad \left| \uv \right|^2
  -\grad V = -g \jtv \; .
\ee Borrowing the identity $\frac{1}{2} \grad \left| \uv \right|^2
= (\uv \cdot \grad) \uv + \uv \times (\grad \times \uv)$, we get,
\be
  \frac{\pd \uv}{\pd t} + (\uv \cdot \grad) \uv
  = -\grad V - \uv \times \o - g \jtv \; ,
  \label{eq:NSE}
\ee 
where $\o \equiv \grad \times \uv$ is the vorticity. This
result reproduces the general NS equation with arbitrary
conservative forces ($\grad V$). The potential $V$ can be
associated with some known forces, for example, ${P}/{\rho}$,
${(G m)}/{r}$ and $\eta(\grad \cdot \uv)$. Here, $P, \rho, G,
\nu + \eta$ denote pressure, density, gravitational constant and
viscosity as well. 

\section{Interaction between soliton and fluid}

In this section we describe an idea to apply the theory described in 
the preceeding section. We give an example on applying the theory 
to provide a consistent way for the interaction between soliton 
and fluid system. Soliton is a 
pulse-like nonlinear wave which forms a collision with similar
pulse having unchanged shape and speed \cite{scott}. The wave
equations that exhibit soliton are the KdV equation, the Nonlinear
Schrodinger equation, the Sine-Gordon equation, Nonlinear
Klein-Gordon equation, the Born-Infeld equation, the Burger
equation and the Boussiness equation. Considering the Nonlinear
Klein-Gordon as follow:
\be
 \frac{\pd^2 \phi}{\pd t^2} - \frac{\pd^2 \phi}{\pd x^2}
 - m^2\phi + \frac{\lambda}{3!} \phi^3=0 \; .
 \label{eq:sg1}
\ee 
The equation is a continuum version of that
describes a propagation of molecular vibration (vibron) in
$\alpha-$helical protein \cite{takeno}. The vibration excitation
in the $\alpha-$helix protein propagates from one group to the
next because of the dipole-dipole interaction between the group.
The wave is called the Davidov soliton \cite{takeno}. Davydov has
shown that in $\alpha-$helical protein soliton can be formed by
coupling the propagation of amide$-I$ vibrations with longitudinal
phonons along spines and that such entities are responsible for
mechanism of energy transfer in biological system \cite{takeno}.
If $\alpha-$helical protein immersed in Bio-fluid, then the
phenomenon can be described by the interaction of soliton with
fluid system. In standard technique in fluid dynamics, the problem
will be done by solving of the Navier-Stokes equation and
nonlinear Klein-Gordon simultaneously. \\

In our current approach the problem is treated as follow. First, 
let us rewrite Eq. (\ref{eq:sg1}) into four vector formalism,
\be
 \pd_\mu \pd^\mu \phi - m^2\phi + \frac{\lambda}{3} \phi^3=0 \; .
 \label{eq:sg2}
\ee 
Using the Euler-Lagrange equation, the lagrangian density is,
\be
 \l = \frac{1}{2}(\pd_\mu \phi)(\pd^\mu \phi) + \frac{m^2}{2!} \phi^2
 - \frac{\lambda}{4!} \phi^4 \; .
 \label{eq:lsg2}
\ee 
In order to couple this lagrangian with the Navier-Stoke lagrangian in 
Eq. (\ref{eq:lagrange1}),
it is sufficient to replace the covariant derivative in Eq.
(\ref{eq:lsg2}) \cite{aslth},
\be
  \D_\mu \phi =(\pd_\mu + ig \A_\mu)\phi \; . 
  \label{eq:covarr}
\ee 
The covariant derivative is invariant under local gauge
transformation\cite{muta}. Then the
interaction between soliton and fluid system obeys the lagrangian,
\be
 \l = -\frac{1}{4}\F_{\mu\nu}\F^{\mu\nu} +
 \frac{1}{2}(\D_\mu \phi)(\D^\mu \phi) + \frac{m^2}{2} \phi^2
 - \frac{\lambda}{4!} \phi^4 \; .
 \label{eq:interlsg}
\ee \\

One interesting case is when we consider a static condition,
\ie $\pd_t f=0$ with $f$ is an arbitrary functions. Substituting
$\A_\mu=(\Phi,-\uv)$ into Eq. (\ref{eq:interlsg}) then the
Lagrange density becomes,
\be
 \l = -\frac{1}{2}(\nabla\times \uv)^2 + \frac{1}{2} |(\nabla -i g \uv)\phi |^2
 +\frac{m^2}{2!} \phi^2 -\frac{\lambda}{4!} \phi^4 \; .
 \label{eq:gl1}
\ee 

\begin{figure}
\begin{center}
 \centering \includegraphics[width=10cm]{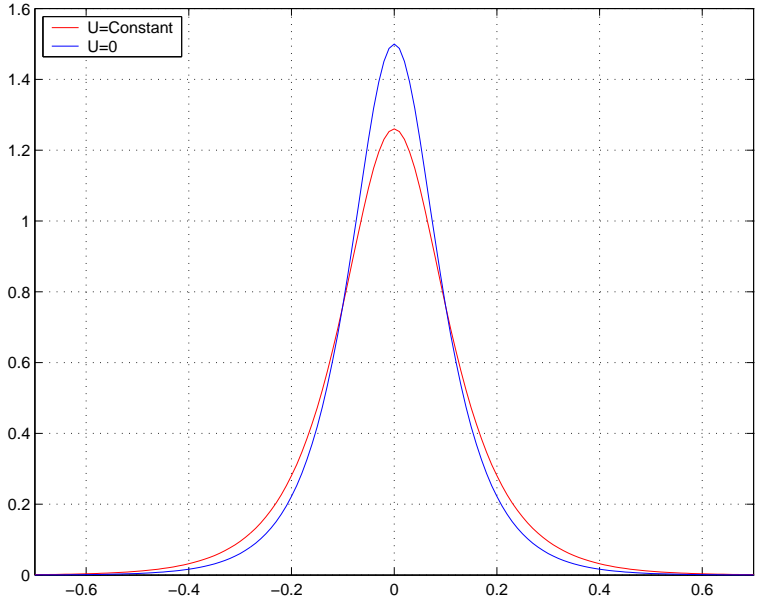}
 \caption{Single soliton solution of the nonlinear Klein-Gordon equation.}
 \label{fig:sghomogen}
\end{center}
\end{figure}

The lagrangian is nothing else similar with the Ginzburg-Landau free energy
lagrangian that is widely used in superconductor theory
\cite{binney}. We have seen that the phenomenon of
$\alpha-$helical protein immersed in fluid similar with quantum
electrodynamics for boson particle, while for static case it is
similar with the Ginzburg-Landau model for superconductor. \\

In order to perform an explicit calculation, suppose we have 
one-dimensional velocity 
in $x$ direction $\uv =(u(x),0,0)$ and $\phi=\phi(x)$. Then the
lagrangian in Eq.(\ref{eq:gl1}) reads,
\be
 \l = \frac{1}{2}\phi_x^2 - \frac{1}{2}g^2u^2 \phi^2
 +\frac{m^2}{2!} \phi^2 -\frac{\lambda}{4!} \phi^4 \; .
 \label{eq:gl2}
\ee 
Substituting it into Euler-lagrangian equation we arrive at,
\be
 \frac{d^2 \phi}{d x^2}  - \gamma(x)\phi + \frac{\lambda}{3!}\phi^3 =0 \; ,
 \label{eq:pSG2}
\ee 
where $\gamma(x)=m^2-g^2u(x)^2 $. The equation is called the
variable coefficient of nonlinear Klein-Gordon equation. \\

Further, we can consider a special case when the fluid velocity is constant, 
$u(x)=U$, to obtain 
\be
 \frac{d^2 \phi}{d x^2}  - \gamma \phi
 +\alpha\phi^3 =0 \; ,
 \label{eq:pSG3}
\ee 
with $\gamma=m^2-g^2U^2$ and $\alpha= {\lambda}/{3!}$. To 
solve the equation, we can use a mathematical trick as follows. First
multiply it by ${d \phi}/{d x}$,
\be
 \frac{d \phi}{d x}\frac{d^2 \phi}{d x^2}  - \gamma \phi \frac{d \phi}{d x}
 + \alpha\phi^3 \frac{d \phi}{d x} =0 \; ,
 \label{eq:pSG4}
\ee 
then integrating out over $x$ and putting the 
integration constant as zero due to integrable condition 
$\lim_{x \rightarrow \pm \infty} \phi=0$. Finally we obtain,
\be
 \left( \frac{d \phi}{d x} \right)^2  - \gamma \phi^2 + \frac{\alpha}{2}\phi^4 =0 \; ,
 \label{eq:pSG5}
\ee 
and it can be rewritten further as,
\be
 \int \frac{d \phi}{\phi (\delta^2 - \phi^2)^{\frac{1}{2}}} = \int \sqrt{\frac{\alpha}{2}} d x \; ,
 \label{eq:pSG6}
\ee 
where $\delta^2 = {2\gamma}/{\alpha}$. Integration of the
left hand side and solving the equation for $\phi$ provide the result, 
\bear
  \phi &=& \frac{2 \delta e^{-\sqrt{\frac{\alpha}{2}} \delta x}}{1+ e^{-2\sqrt{\frac{\alpha}{2}}\delta
  x}}  = \frac{2 \delta}{e^{\sqrt{\frac{\alpha}{2}} \delta x} + e^{-\sqrt{\frac{\alpha}{2}}\delta x}}\nonumber \\
  &=& \frac{\delta}{\cosh (\sqrt{\frac{\alpha}{2}}\delta x)}
0  =\delta \mathrm{sech} (\sqrt{\frac{\alpha}{2}} \delta x) \; .
 \label{eq:pSG8}
\eear 
Thus, the solution for a homogeneous nonlinear Klein - Gordon
equation is,
\be
 \phi(x)= A \mathrm{sech} (\Lambda x) \; ,
 \label{eq:pSG9}
\ee 
where $A= {(12 \gamma)}/{\lambda}$ and 
$\Lambda= {(12 \sqrt{3} \gamma)}/{\lambda^{3/2}}$. This result is depicted in
Fig. \ref{fig:sghomogen}. 
The figure shows that the soliton propagation will be damped by
fluid. This theory also can be applied in turbulence phenomenon
\cite{sulaiman}.

\section{Conclusion}

We have shown an analogy between electromagnetics
field and fluid dynamics using the Maxwell-like equation for an
ideal fluid. The results provide a clue that we might be able to
build a gauge invariant lagrangian density, the so-called  
Navier-Stokes lagrangian in term of
scalar and vector potentials $\A_\mu$. Then the Navier-Stokes
equation is obtained as its equation of motion through the
Euler-lagrange principle. The
application of the theory is wide, for instance the interaction
between Davydov soliton with fluid system that can be described by
the lagrangian density which is similar to quantum electrodynamics
for boson particle. In the static condition, the lagrangian
density is similar with the Ginzburg-Landau lagrangian. If the
fluid flow is parallel with soliton propagation we also obtain the
variable coefficient Nonlinear Klein-Gordon equation.  Single
soliton solution has been obtained in term of a second hyperbolic
function. The result showed that the present fluid flow will
give a damping in solitary wave propagation.

\section*{Acknowledgment}

The authors thank Terry Mart, Anto Sulaksono and all of the theoretical
group members (Ketut Saputra, Ardy Mustafa, Handhika, Fahd, Jani, Ayung) 
for so many valuable discussion. This research is partly funded by DIP 
P3-TISDA BPPT and Riset Kompetitif LIPI (fiscal year 2005).


\begin{thebibliography}{99}
\bibitem{kundu} P. Kundu (1996),
    \textit{Fluids Mechanics}, Addison-Wesley, New York.
\bibitem{mulin} T. Mulin (1995), 
    \textit{The Nature of Chaos}, Clarendon Press, Oxford.
\bibitem{sulaiman} A. Sulaiman (2005),  
	\textit{Contruction of The Navier-Stokes Equation using Gauge Field Theory  Approach}, Master Theses at Department of Physics, University of Indonesia.
 \bibitem{aslth} A. Sulaiman and L.T. Handoko (2005), Gauge field theory approach to construct the Navier-Stokes equation, \textit{arXiv:physics/0508086}. 
\bibitem{huang2} Huang.K (1992),
    \textit{Quarks, Leptons and Gauge Fields}, Worlds Sceintific, Singapore.
 \bibitem{muta} Muta.T (2000), 
    \textit{Foundation of Quantum Chromodynamics}, Worlds Sceintific, Singapore.
\bibitem{binney} Binney. J.J  et.al. (1995),
   \textit{The Theory of Critical Phenomena}, Clarendon press, Oxford.
\bibitem{takeno} Takeno, S (1987), Vibron Soliton and Coherent
Polarization, \textit{Collected paper Dedicated to prof K Tomita},
Editor:Takeno.S et al , Kyoto University Press. Kyoto.
 \bibitem{scott} A. Scott, et al (1973), Soliton: A New Concepts
 in Applied Science,\textit{Proceeding of the IEEE}, \textbf{61}, 1443-1464.
\end{thebibliography}
\end{document}